# 60-cm-long In-vacuum Zerodur Optical Cavity for Optical Lattice Atom Interferometry

Philip Wondrak[1], Paola Luna[1*], Kenneth Nakasone[1*], Andrei Zhukov[1*], Michael Grisham[2], Garrett D. Cole[1], Po-han Chen[1], Cristian D. Panda[1]

[1]*Wyant College of Optical Sciences, The University of Arizona, Tucson, AZ 85721, USA*

[2]*Stable Laser Systems, Boulder, CO 80301, USA*

[*]*The authors contributed equally to this work.*

Optical cavities are widely used in atomic physics experiments to control the spatial geometry and spectral properties of atom-light interactions. Recently, an optical lattice generated by the pristine mode of an optical cavity was utilized to hold and extend the quantum coherence of atomic spatial superpositions. Here, we describe an in-vacuum optical cavity designed to enhance the spatial and temporal coherence of optical lattice atom interferometers to one millimeter wavepacket separations and one minute measurement durations. The cavity uses a rigid Zerodur spacer to suppress the negative effects of thermal drift and mechanical-acoustic vibrations on the atomic motional state. The 60 cm-long cavity spacer provides ample optical access for the laser beams and imaging optics necessary for the preparation, quantum control, and read-out of the atomic state, as well as room for various source mass configurations for localized gravitational measurements. An integrated differential pumping structure enables compatibility with ultrahigh-vacuum operation and optical interferometry allows for in-situ vibration monitoring. The performance of the optical cavity is methodically characterized: we measure mirror vibration noise with amplitude below $5\,nm/\sqrt{Hz}$ in a wide frequency range of 2 Hz - 100 kHz and a slow 15 MHz thermal drift of the cavity resonance over 47 hours, which can be readily stabilized with an integrated piezo. The optical cavity supports both 852-nm light for Raman transitions and 1064-nm light for the optical lattice, with a free spectral range matched to an integer divisor of the cesium ground-state hyperfine splitting.

# Introduction

Optical cavities have numerous applications in atomic and optical physics experiments, including cavity quantum electrodynamics[1,2], frequency metrology[3,4], and atomic timekeeping[5,6]. In atom interferometry, optical cavities can provide power enhancement, atom localization, and spatial mode filtering of the laser fields used for atomic manipulation[7,8], while cavity-mediated interactions have also enabled entanglement-enhanced measurements[9].

More recently, optical cavities have played an important role in extending the interrogation times achievable in atom interferometry. In contrast to conventional atom interferometers, in which atoms undergo free fall, optical lattice atom interferometers (OLAIs) suspend atomic spatial superposition states in a standing wave optical lattice. Using an optical lattice filtered by an optical cavity, the quantum coherence of spatial superpositions of cesium atoms have been maintained for as long as 70 seconds[10]. This technique has also enabled measurements of the gravitational attraction of a miniature source mass, demonstrating its potential for precision measurements of gravity and searches for new fundamental interactions[11].

Present OLAI performance depends critically on the stability and spatial uniformity of the optical hold potential, where the optical cavity provides spatial filtering of small scale imperfections in the optical lattice field, such as speckle and wavefront distorsions. However, fluctuations in the large-scale global properties of the optical lattice mode, including position, intensity, and spatial structure, can introduce phase noise, motional excitation, and decoherence of the atomic spatial superposition[10]. These global optical properties of the cavity mode are heavily dependent on the mechanical and thermal stability of the spacer and the cavity mirrors.

The stability of the optical mode is exquisitely well-controlled in the ultrastable optical reference cavities used in optical frequency metrology. These cavities typically use low-expansion materials (e.g. ULE, Zerodur, silicon crystal) for their spacers and mechanically optimized geometries that suppress sensitivity to temperature fluctuations and external vibrations[5,12–15]. Integrating these design principles with OLAI experiments requires consideration of additional requirements. As examples, the cavity must accommodate space for magnetic field-generating coils and integration of gravitational source masses, and optical access for the laser beams used for atomic cooling, quantum state preparation and manipulation, and imaging. The cavity must also operate within a non-magnetic, ultra high-vacuum environment.

Here, we present the design, construction, and characterization of an in-vacuum, 60-cm-long, optical cavity for OLAI. The cavity incorporates a vibrationally-stable Zerodur spacer assembly that addresses decoherence of atomic spatial superpositions due to transverse motion of the optical lattice, which limited spatial and temporal coherence in previous optical lattice atom interferometers[10]. The Zerodur spacer assembly features a low coefficient of thermal expansion, large openings for optical access, and a differential pumping structure that isolates the interferometer vacuum section from the atomic source. The cavity mode supports both 852-nm light for Raman atomic beamsplitter transitions and 1064-nm light for the optical lattice for atom interferometry. The cavity length is tuned so that the free spectral range is an integer divisor of the cesium ground-state hyperfine splitting. Thereby, the cavity simultaneously enhances the two frequency tones needed for Raman transitions. Construction uses non-magnetic, vacuum-compatible materials, and installation is facilitated by a custom-made elevator. The resulting instrument provides a stable optical platform with flexible experimental access for precision OLAI measurements and future quantum sensing applications.

# Design and Implementation

The cavity is designed to operate within a high precision OLAI laboratory setup, where integration and modularity are important principles. Integration of a scaled-down version of the optical cavity into a fieldable compact mobile gravimeter setup for quantum inertial sensing applications is possible in the future, although not the primary goal here.

The optical cavity length, $L_c$ = 60.33 cm, was chosen to accommodate 10 cm-scale gravitational source masses for use in laboratory gravitational measurements that range from measuring Newton's gravitational constant[16,17] to testing dark energy models[11,18]. Figure 2 illustrates a possible source mass geometry, a hollow cylinder placed outside of the vacuum chamber. The optical cavity length accommodates various differential gravitational measurement schemes designed to reduce systematic effects due to background fields and source mass non-uniformity[11]. Differential measurements with the atoms located at symmetric positions around the cavity waist cancel out interferometer phases from differential lightshifts due to laser divergence.

In addition, these systematic effects are minimized by holding the atoms near the cavity waist[11]. For simplicity, the cavity is symmetric, i.e. uses two mirrors of equal radii of curvature, so that the cavity waist position matches the geometrical center of the spacer. The atoms are loaded into the cavity mode from a magneto-optical trap (MOT) located 24 cm below the cavity waist, at the center of the MOT section of the vacuum system (Fig. 2), and can be translated to any location along the cavity axis by using an atom elevator.

A longer cavity has additional advantages: the cavity waist scales approximately as $L_c^{1/4}$. A large waist provides flat, uniform optical wavefronts that mitigate the detrimental effects of transverse motion of the optical lattice on the coherence of the trapped atomic spatial superposition[10]. However, low expansion spacers are rarely manufactured with lengths greater than ~30 cm and maintaining good mechanical stability and optical alignment in long cavities can be challenging[19,20]. Our cavity integrates three separate spacers to accommodate this length limitation (see section Manufacturing and Assembly).

The cavity length was carefully adjusted to achieve a free spectral range of $2\pi \times$ FSR= $2\pi \times$ 248.448 ± 0.006 MHz, which is an excellent match to the cesium hyperfine frequency ($2\pi \times$ 9192.631770 MHz) divided by a factor of 37. This ensures that the cavity is simultaneously resonant with the laser frequencies that drive Raman transitions between the cesium hyperfine states.

The mirror radius of curvature, R = 15 m, is among the longest commercially available for superpolished surfaces, providing large waists of 756 μm and 844 μm at the wavelengths of the Raman laser ($\lambda_{Raman} = 852\, nm$) and optical lattice laser ($\lambda_{lattice} = 1064\, nm$), respectively. This cavity geometry results in a transverse mode spacing of 22.5 MHz, several times larger than the combined linewidth of the cavity and cavity-coupled lasers. Components of the input laser beam that overlap with high-order spatial cavity modes are therefore filtered out, resulting in a clean, near-Gaussian, intra-cavity mode that minimizes quantum decoherence[10].

Typical multi-layer dielectric coatings provide amazing flexibility for accurate tuning of mirror reflectivities[21,22]. In our cavity, mirror reflectivities at $\lambda_{Raman} = 852$ nm and $\lambda_{lattice} = 1064$ nm are 0.945 and 0.981, respectively. These result in linewidths of 4.48 MHz and 1.52 MHz at the two wavelengths, corresponding to finesses of 55 and 164, respectively.

The cavity linewidth of 4.48 MHz at $\lambda_{Raman} = 852\, nm$ is sufficient to accept near-resonant cavity coupling of the Raman laser for atomic velocities in a range up to 2 m/s (corresponding to a Doppler detuning of 4.7 MHz), while still providing an intra-cavity power enhancement factor of 18 and good spatial mode filtering. A narrower cavity linewidth of 1.52 MHz at $\lambda_{lattice} = 1064\, nm$ represents a compromise between spatial mode filtering of the optical lattice laser and suppression of laser frequency-to-amplitude noise conversion. The latter is important because optical lattice intensity fluctuations can induce parametric heating of the trapped atoms, thereby limiting their lifetime in the optical lattice[23,24].

## Manufacturing and Assembly

The geometry of the optical cavity is shown in Figure 1. The cavity spacer assembly consists of three connected Zerodur pieces: a 313 mm interferometer spacer, a 200 mm MOT spacer, and a shorter 88 mm top spacer. The top spacer is carefully ground down to a specified length that ensures that the total cavity length is an exact half-integer multiple of the cesium hyperfine wavelength (32.61225572 mm), as required for simultaneous resonance of the two Raman tones. The three spacers are mechanically joined with dowel pins and 316L stainless steel screws. The mirrors are attached to the ends of the spacer assembly using aluminum clamps and springs. Cavity alignment is fine tuned using 12 μm thick Invar 36 metal shims between the spacer pieces. The cavity is attached to the vacuum chamber at two points: a rigid supporting connection uses a stainless steel plate (modified Kimball MCF450-GrvGrb-H-Blank) mounted to the spherical octagon MOT chamber and a flexible connection uses four Viton spheres that are squeezed between the cavity spacer and a custom plate holder mounted into a Kimball 4.5” ConFlat close coupler (Model MCF450-ClsCplr-E2).

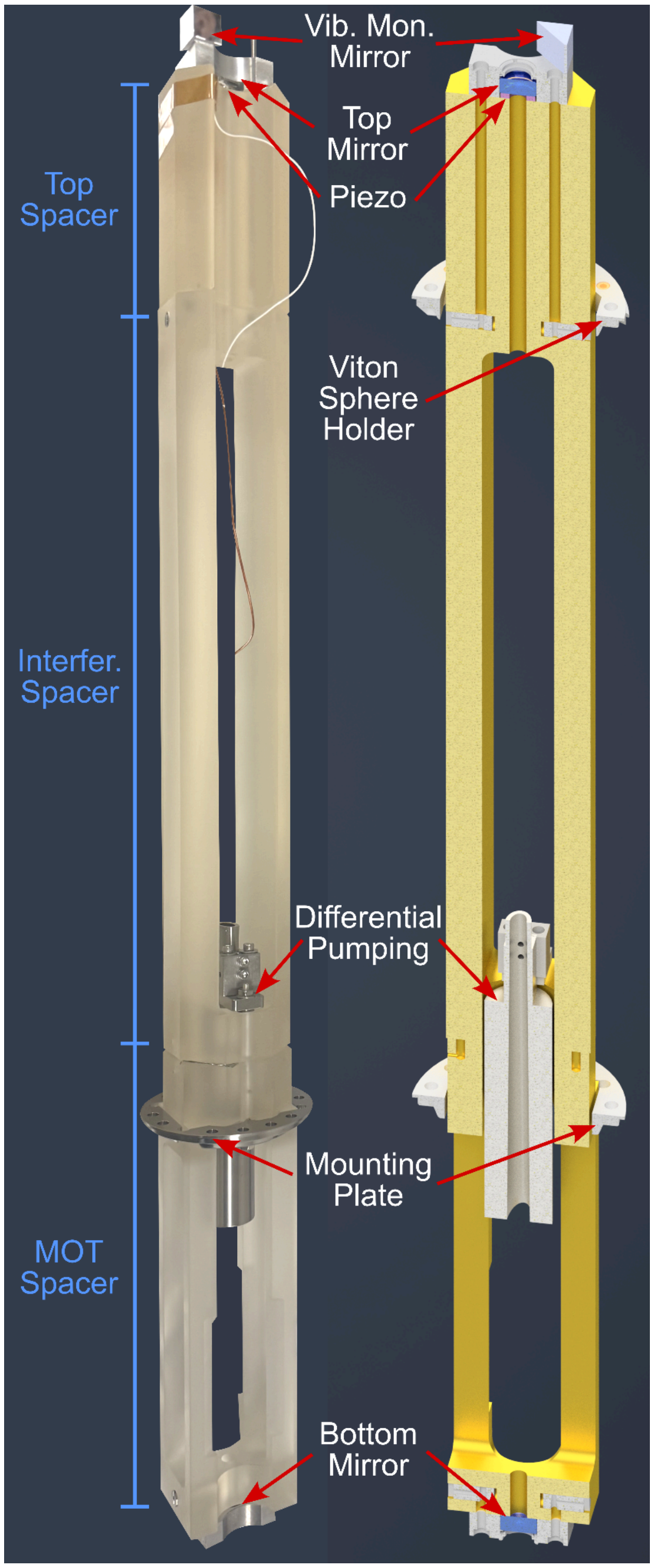

**Figure 1. Image (left) and CAD cross-section (right) of the assembled cavity.** Three custom-machined Zerodur pieces are clamped together using screws. They are designated, from top to bottom: top spacer, interferometer spacer and MOT spacer. The two optical cavity mirrors are mounted at the ends of the complete spacer assembly. The “MOT spacer” has a rectangular profile

with cutouts for multiple laser beams that are used for atomic physics processes, including MOT generation, laser cooling, and excess atom removal. The interferometer spacer has a cutout that allows for the imaging of the atom sample at multiple locations along the cavity axis and for the insertion of a miniaturized, cm-scale source mass. A 316 stainless steel mounting plate is used to attach the cavity to the conflat flange of the vacuum chamber. An additional connection point to the vacuum chamber is made using flexible vacuum-compatible Viton spheres. A titanium differential pumping section isolates the vacuum of the interferometer region from the MOT region, where the pressure is limited by the cesium atomic background.

The MOT section of the cavity has a rectangular profile with openings on all four sides. This design offers space for the in-vacuum coils used for atomic evaporative cooling, [25] and ample optical access for all the laser and microwave fields used for atom cooling, state preparation and imaging, including the six laser beams that form the cesium magneto-optical trap[26]. The interferometer section has a side opening that allows for atom removal with a resonant laser beam, imaging, and insertion of cm-sized source masses for short-range gravity measurements. Additional circular cutouts in the MOT spacer allow for passage of resonant laser beams at small angles with respect to the cavity axis that address atoms loaded at any position along the optical cavity mode.

The MOT and interferometer sections are connected with a differential pumping section, which is designed to minimize residual vacuum pressure in the interferometer region due to contaminants in the MOT region. Contaminants include cesium atoms that are not removed by the MOT vacuum pump. The design geometry was optimized and tested by running molecular flow simulations of the vacuum system geometry (MolFlow v 2.9). Simulations predict that the pressure due to contaminants in the MOT region (such as thermal cesium atoms) is reduced 20 fold in the interferometer region in the final OLAI configuration.

The differential pumping section comprises a 12.7 cm long titanium cylinder with an elliptical inner bore hole. The dimensions of the bore are optimized to reduce the vacuum conductivity between the two sections, while allowing optical access for the optical cavity mode and the resonant atom removal laser beam. This titanium cylinder is mounted to the interior of the Zerodur spacer using vented non-magnetic screws.

Monitoring of cavity vibrations in the plane transverse to the cavity axis is achieved with an optical Michelson interferometer. To obtain a reference for measuring cavity displacement, we epoxied a “vibration monitoring mirror” (using TorrSeal) onto the top of the cavity. The mirror has two orthogonal planar surfaces that can be used to monitor vibrations along both spatial axes. The mirror reflects an incident beam back along the measurement arm of the Michelson interferometer (see section Testing Methods and Results).

# Setup for In-vacuum Cavity Testing

The vacuum setup we used for cavity testing required capabilities to characterize the following parameters: (1) optical properties of the cavity including free spectral range, transverse mode spacing, long-term thermal drift in the cavity resonance, and cavity linewidth, (2) vacuum performance, and (3) vibrational response of the cavity to environmental mechanical and acoustic noise.

A diagram of the test setup is shown in Figure 2. The vacuum assembly was built around a Kimball 8” spherical-square vacuum chamber octagon (MCF800-SphSq-G2E4C4), which we refer to as the MOT section. The interferometer section of the vacuum system was mounted on top of the MOT section and consisted of four vertical stacked 4.5” vacuum parts (Kimball close coupler (MCF450-ClsCplr-E2), 3.75” long nipple, a vacuum tee, and a top cross).

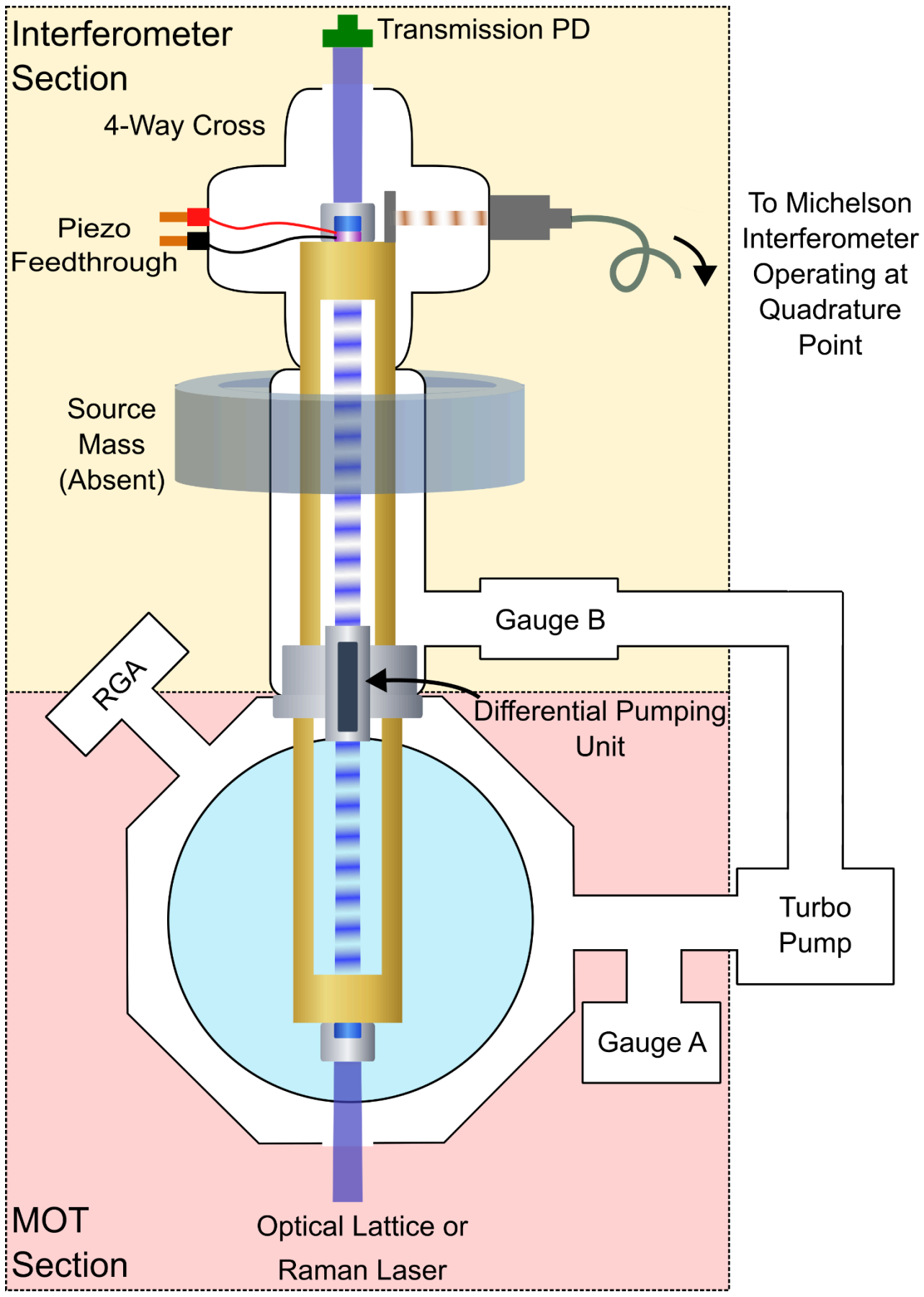

**Figure 2. Configuration for in-vacuum testing.** The cavity (yellow) was mounted inside a conflat vacuum system with two regions: a MOT section and an interferometer section. A turbo pump evacuated both sections. Gauges A and B measured the pressures in the MOT and interferometer sections, respectively. A residual gas analyzer (RGA) was used to characterize the vacuum quality. Optical tests were performed using a laser coupled to the cavity (vertical blue beam), with throughput measured on a transmission photodetector (PD). Vibration tests used a separate laser beam (horizontal bourbon-color striped beam) launched from a fiber towards the mirror attached to the top spacer (vibration monitoring mirror in Fig. 1) and retroreflected back into the fiber. We used a piezo to scan the cavity length. The source mass pictured was absent from this test setup, but illustrates the utility of a 60-cm long cavity for future OLAI gravitational measurements.

An Agilent TPS-Compact turbo pump was connected in two places: to the bottom MOT section and to the top interferometer section. Two ion gauges were used to measure the pressures in the two vacuum sections, Gauge A (Inficon MPG400 DN 40 CF-F Vacuum Gauge) monitored the pressure in the MOT section and Gauge B (Varian LR88590 L8350-301) monitored the pressure in the interferometer section. To enable optical testing, the vacuum system was fitted with viewports for the optical test beams illustrated in Figure 2.

Installation of the cavity into the vacuum system occurred in several stages. The cavity was lowered through the top flange opening of the MOT section vacuum chamber using a guiding jig and rigidly attached to the top flange by tightening the mounting plate screws onto the chamber grabber grooves with commercially available mounting hardware (Kimball Physics). We then used an elevator to lower the interferometer section vacuum parts around the cavity, completing the cavity enclosure. The elevator was custom made from extruded aluminum rails mounted to the legs of a consumer-grade standing desk. The elevator included alignment plates to translate vacuum parts smoothly during installation, thereby minimizing the risk of collision between the metal vacuum chamber parts and the glass cavity spacer.

# Testing Methods and Results

## Vacuum Testing

We used the system shown in Figure 2 to test the vacuum compatibility of the cavity. The turbo pumping configuration described above was only able to achieve pressures in the $10^{-8}$ torr range (measured by gauges A and B), significantly above those necessary for an atom interferometer vacuum system, which uses combined ion-getter pumps to reach $10^{-11}$ torr. The entire system was heated to 110 $^{o}$C, which tested the resilience of the cavity assembly to the bakeout process needed to achieve UHV pressures in the OLAI system.

A residual gas analyzer (RGA) attached to the MOT chamber (Fig. 2) was used to measure the partial pressures before, during, and after bake out. The RGA scans show that primary residual pressure peaks come from water and air leaks through the turbo pump.

## Optical Testing

We tested the optical performance of the cavity using two lasers: the Raman laser (852 nm) and the optical lattice laser (1064 nm). The Raman laser (Toptica DL Pro) was offset locked[27,28] via piezo feedback to a reference laser stabilized to the D2-line of a Cesium reference cell via modulation transfer spectroscopy (MTS)[29]. The offset frequency of 3*($2\pi \times 14$) GHz was given by the third-order sideband modulation frequency of a fiber electro-optic modulator (Exail NIR-MPX800-LN-20) driven with a MW function generator (Windfreak SythHD) at $2\pi \times 14$ GHz. This offset was chosen to detune the Raman laser sufficiently from the atomic resonance to avoid photon scattering during OLAI experiments. The optical lattice laser was derived from an amplified (Toptica ALS-IR-1064) ECDL (Rio Orion-1064) that was kept free-running for the tests described here. We coupled light into the cavity with a 6.2 mm focal length asphere collimator from a single mode optical fiber (Thorlabs PM 780).

We characterized the cavity linewidth at the Raman laser wavelength (852 nm) by sweeping the laser frequency over a range of $2\pi \times$ 600 MHz and measuring the transmitted cavity power on a linear photodiode, as shown in Figure 3. The piezo was grounded for this measurement. The figure shows cavity resonances corresponding to the lowest transverse modes up to order two over two FSRs. We fit four Lorentzian lineshapes to the zeroth, first, and two non-degenerate second-order transverse HG modes. Table I shows the results averaged over five independent scans. The cavity linewidth (Gaussian one-sigma width) at 852 nm is extracted from the fits to the $HG_{00}$ modes. The cavity transverse mode spacing is extracted from the frequency difference between the zeroth order and the average of the two non-degenerate second-order transverse HG modes.

Similarly, the cavity linewidth at 1064 nm was measured by modulating the piezo voltage of the Zerodur cavity while the optical lattice laser frequency was left unmodulated. The frequency reference for this measurement was the transverse mode spacing measured above.

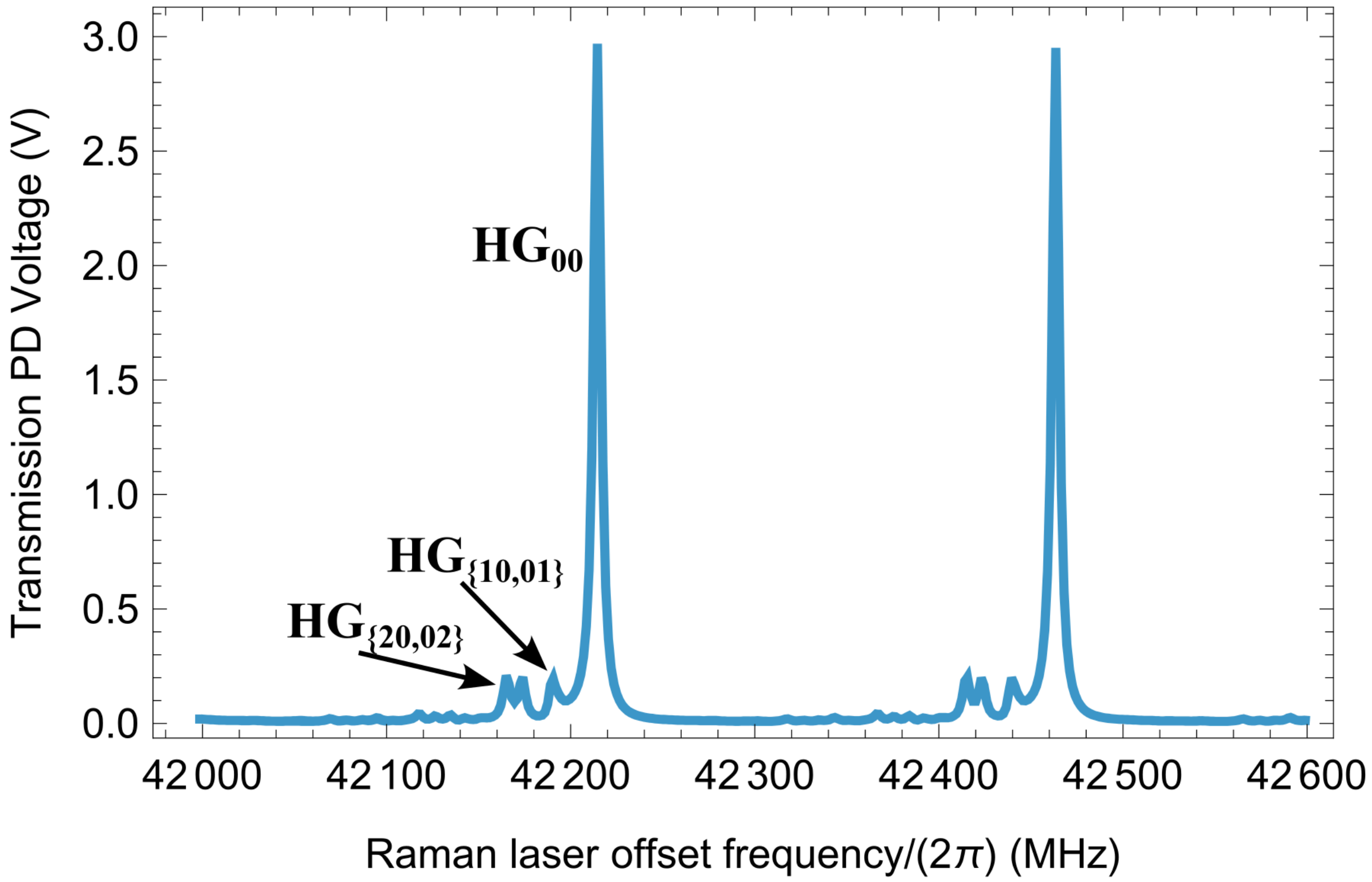

**Figure 3. Measured cavity spectrum at 852 nm.** Resonances corresponding to the lowest order transverse cavity modes are visible. The two sets of cavity modes are separated by one FSR.

The cavity FSR was verified by adding sidebands on the locked Raman laser light using a separate electro-optic modulator (Thorlabs LNY8510A) driven with a microwave generator (Valon Technology 5015 Frequency Synthesizer) at a modulation frequency set near the hyperfine cesium resonance ($\omega_{HF}$=$2\pi \times$ 9.192631770 GHz). The absolute frequency of the microwave generator was referenced to a 10 MHz stable oscillator (Wenzel Premium 10 MHz -

SC Blue Top Ultra Low Noise Osc. 501-27503). Our cavity is designed so the cavity FSR is an exact integer divisor of $\omega_{HF}$ ($\omega_{HF}$/37). When this condition is perfectly achieved, the first order sidebands overlap with the carrier spectral peak as the cavity is swept through resonance. To check the offset from this design condition, we recorded two modulation frequencies at which the positive and negative sidebands have identical frequency offsets (equal detuning) from the cavity resonance. Taking the mean of these two frequencies and dividing by 37 yields a measurement of the cavity FSR. Using this method, we determine that the cavity FSR=248.446 ± 0.006 MHz, only 3 kHz away from the ideal FSR. This corresponds to a cavity length of 60.3 cm that is matched to within 400 nm. All the other tested optical parameters are in very good agreement with the design.

**Table I. Optical properties.** The table includes design and measured parameters. An asterisk indicates that the measured linewidths are combined laser and cavity linewidths, rather than only the cavity linewidth shown under the Design column.

| **Parameter** | **Design** | **Measured (MHz)** | **Measurement Uncertainty (MHz)** |
|---|---|---|---|
| FSR | 248.449 | 248.446 | 0.006 |
| Linewidth (852 nm) | 4.48 | 4.7* | 0.1 |
| Linewidth (1064 nm) | 1.52 | 2.0* | 0.1 |
| Transverse mode spacing | 22.5 | 22.3 | 0.3 |

The long-term drift of the cavity resonant frequency was monitored with a similar test configuration to that which used the Raman laser to measure the cavity linewidth. We grounded the piezo and scanned the stabilized Raman laser in a range of ±100 MHz around the $HG_{00}$ cavity resonance. We fit the Lorentzian resonance peak and monitored the drift in the center frequency over a period of 47 hours (Fig. 4). The temperature of the vacuum chamber was monitored over the same time period using five separate thermistors evenly distributed on the exterior surface of the vacuum chamber. A drift of ±10 MHz from the initial resonant frequency is observed over the 47 hour period. We expect and observe a time delay of ~1-2 h for the cavity temperature to approach thermal equilibrium with the vacuum chamber. After accounting for this, we observe a strong correlation between the cavity resonant frequency and measured temperature. The temperature to frequency conversion shown in Figure 4 yields a cavity linear coefficient of thermal expansion (CTE) of $1.1 \cdot 10^{-7}$.

One of the main usability limitations of previous OLAIs[10] was drift in the cavity resonance of up to 2-3 GHz over a daily run due to changes in vacuum chamber length with temperature. Correcting for this with the piezo is an imperfect solution, since piezo non-linearities introduce tilting of the mirror that is dependent on the piezo voltage. These tilts significantly shift the cavity mode spatially, affecting its angular stability and alignment with gravity.

Here, we reduce the drifts in cavity length with temperature more than 200-fold due to the lower CTE of the Zerodur spacer compared to the steel vacuum chamber, despite an increase in the cavity length by a factor of two. The 15 MHz drift can be readily corrected with feedback on the integrated piezo, whose range of motion can now be reduced by the same factor.

In the future, stabilizing the environmental temperature of the cavity enclosure to the zero-crossing of the CTE for the assembly can further increase the long-term absolute stability of the cavity frequency resonances. Nevertheless, the system is much more robust to changes in the environmental temperature, which represents a critical step towards developing cavity-based OLAIs for operation in challenging field environments for quantum inertial sensing applications[30,31].

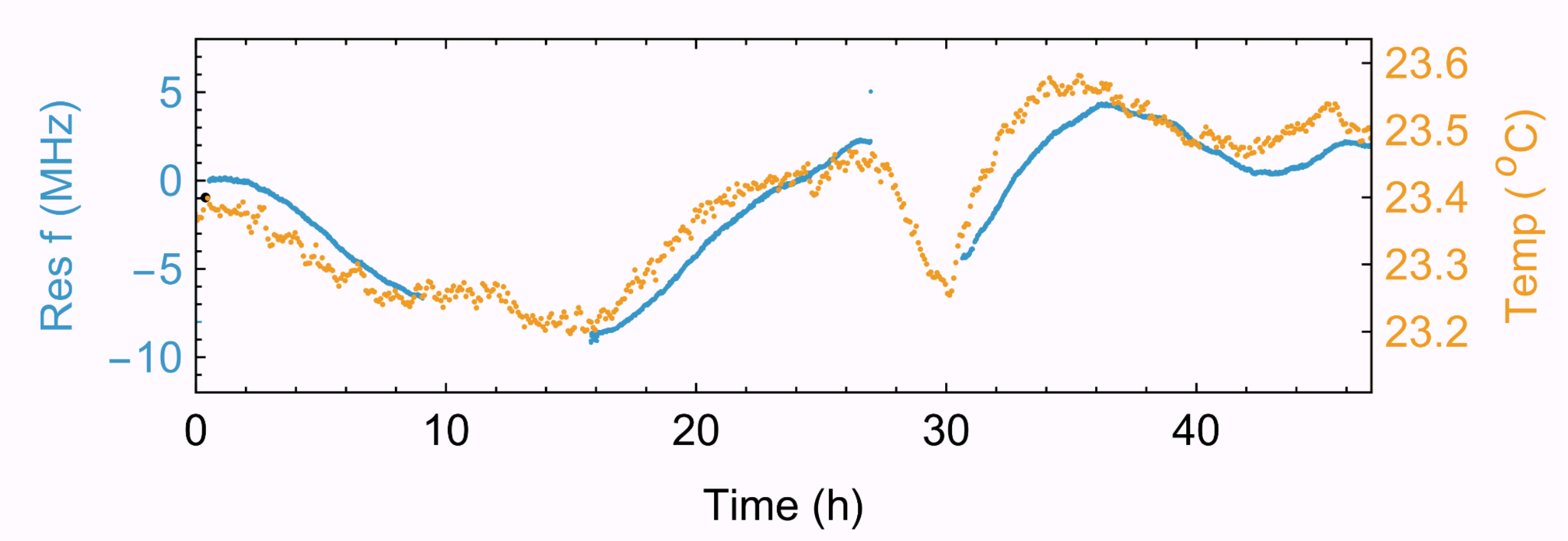

**Figure 4. Measured drift in cavity resonance over 47 hours.** Blue data points show the drift in the cavity resonance. Orange data points show the drift in the mean temperature recorded using 5 thermistors attached to the exterior of the vacuum chamber. Conversion between the temperature axis and the resonant frequency axis yields a linear CTE of $1.1 \cdot 10^{-7}$. We observe a strong correlation between the cavity temperature and resonant frequency.

## Vibrational testing

Michelson interferometers are commonly used in various configurations to characterize vibrations[32–34]. Similarly, we used a Michelson interferometer with a separate beam path from the science laser as a vibrometer to characterize the vibration sensitivity of the cavity-vacuum setup. In the measurement arm of the interferometer, the laser was coupled into a single mode fiber. The fiber output was delivered from an adjustable collimator mounted directly to the vacuum chamber, onto the vibration monitoring mirror glued to the top of the cavity (Fig. 1). The fiber collimator was adjusted to maximize coupling of the retroreflected beam back into the fiber to complete the measurement arm. The Michelson interferometer measurement and reference arms were overlapped at the beam splitter and the output was monitored with a photodiode. This configuration thereby measures the relative vibrations of the cavity with respect to the four-way cross (Fig. 2).

We move a translation stage in the reference arm to gauge the difference between the maximum and minimum photodiode voltages, $V_{amp}$, corresponding to constructive and destructive interference, respectively. When the average photodiode voltage settles halfway between the maximum and minimum, we record a time series of 2.4 seconds, during which the photodiode voltage remains near the midpoint. This corresponds to the quadrature point of the interferometer, where total optical path difference (OPD) between the two arms is $\lambda/4$. During this measurement, the voltage does not exceed the maximum and minimum points, corresponding to fluctuations in absolute mirror displacement below ($\lambda$/4)/2. This eliminates the need to unwrap the interferometer phase during data processing.

We process the data by first removing the DC voltage offset, and obtain $V_{ac}$. We then convert $V_{ac}$ to displacement via the equation

$$\Delta x = \frac{\lambda}{4\pi} \arcsin\left(\frac{V_{ac}}{V_{amp}}\right).$$

The displacement trace was then converted to an amplitude spectral density (ASD) using the “compute_spectrum()” method from the public SpecKit python toolkit.

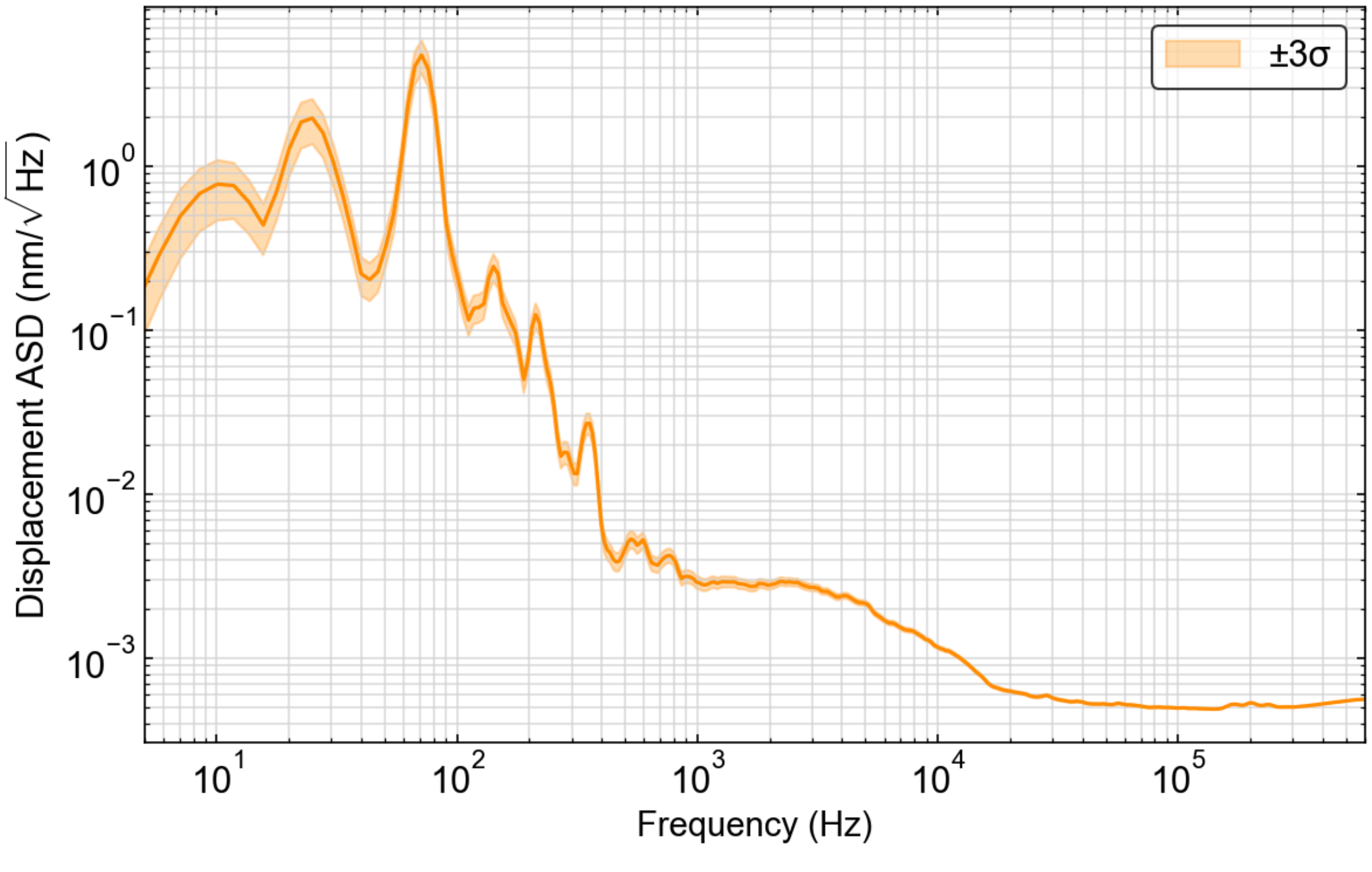

**Figure 5. ASD plot of the top cavity mirror displacement in the direction transverse to the cavity axis.** A Michelson interferometer operating at the quadrature point was used to measure the position of the vibration monitoring mirror with respect to the top cross of the vacuum setup. The spectrum above was derived from a 2.4 second oscilloscope trace. The largest resonance observed in the spectrum resides at 70 Hz and is attributed to the first bending mode of the vacuum setup. Transparent yellow bands correspond to a $\pm 3\sigma$ confidence interval.

The resulting ASD is shown in Figure 5. The largest resonance around 70 Hz, with peak amplitude $5\,nm/\sqrt{Hz}$, likely corresponds to the excitation of the first bending mode of the cavity-vacuum system, which was clamped to the optical table at the bottom and left unsupported at the top for this measurement. Additional resonances are observed in the 5 - 400 Hz band. In separate testing, we correlate these noise peaks with acousto-mechanical noise sources in the laboratory. The experimental table was not floated for any of these tests. Floating the experiment table and reducing environmental sources of mechanical-acoustic noise should further reduce background noise. Low frequency resonances in the cavity and vacuum enclosure system will be further suppressed by clamping sections of the experimental systems.

# Conclusion

We have described the design, construction, and characterization of a 60-cm-long in-vacuum Zerodur optical cavity for optical lattice atom interferometry experiments. The cavity combines improved mechanical and thermal stability with extensive optical access for implementing precise quantum control of ultracold atoms and space for various gravitational source mass geometries. We characterized the optical properties of the cavity at both 852 nm and 1064 nm, finding good agreement with the design parameters. The cavity exhibits long-term frequency stability better than 15 MHz over approximately two days, a substantial improvement over designs that mount the cavity mirrors directly to the steel vacuum chamber. We used a Michelson interferometer to measure a transverse vibrational spectrum with peak amplitudes less than $5\,nm/\sqrt{Hz}$ in a wide frequency range of 2 Hz - 100 kHz.

The improved stability of the optical lattice is expected to reduce motional excitation and decoherence of trapped atomic spatial superpositions, enabling optical lattice atom interferometry with minute-long interrogation times and mm-scale arm separations. The large optical access and stable cavity geometry provides a platform for future measurements of gravitational interactions, including measurements of Newton's gravitational constant and searches for new physics at short distances.

## Acknowledgements

We thank Daniel Caywood, Felipe Guzman, Jose Hernandez, Chang Jin Oh, Jason Jones, Mark Notcutt, Jose Sanjuan, and Dal Wilson for helpful discussions.